\documentclass{ws-procs9x6}

\begin{document}

\title{Long Baseline Neutrino Oscillations:\\
Parameter Degeneracies\\ and\\ JHF/NuMI 
Complementarity\footnote{http://www-sk.icrr.u-tokyo.ac.jp/noon2003/transparencies/11/Parke.pdf}}

\author{Hisakazu Minakata}
\address{Department of Physics, Tokyo Metropolitan University \\
1-1 Minami-Osawa, Hachioji, Tokyo 192-0397, Japan\\
Email: minakata@phys.metro-u.ac.jp}

\author{Hiroshi Nunokawa}
\address{Instituto de F\'{\i}sica Te\'orica,
Universidade Estadual Paulista \\
Rua Pamplona 145, 01405-900 S\~ao Paulo, SP Brazil\\
Email: nunokawa@ift.unesp.br}

\author{Stephen Parke\footnote{\uppercase{P}resenter at 
\uppercase{NOON} 2003.}}
\address{Theoretical Physics Department,
Fermi National Accelerator Laboratory \\
P.O.Box 500, Batavia, IL 60510, USA\\
Email: parke@fnal.gov}

\maketitle

\abstracts{
A summary of the parameter degeneracy issue for long baseline neutrino
oscillations is presented and how a sequence of measurements can be used
to resolve all degeneracies.
Next, a comparison of the JHF and NuMI Off-Axis proposals is made with emphasis
on how both experiments running neutrinos can distinguish between the 
normal and inverted hierarchies provided the E/L of NuMI is less than
or equal to the E/L of JHF.
Due to the space limitations of this proceedings only an executive style
summary can be presented here, but the references and transparencies of the 
talk contain the detailed arguments. 
}
\section{Parameter Degeneracies: Overview}
The probability of $\nu_\mu \rightarrow \nu_e$ depends on
$\theta_{13}$, $\delta_{CP}$, $\theta_{23}$ and sign of $\delta m^2_{31}$.
Untangling the degeneracies associated with these parameters is the subject of
this section, 
see Ref. [\refcite{Burguet-Castell:2002qx}]-[\refcite{Minakata:2002qi}].

\subsection{$\theta_{13}$ and $\delta_{CP}$ Degeneracy}\label{subsec:13delta}
\begin{figure}[hbt]
\centerline{\epsfxsize=4.2in\epsfbox{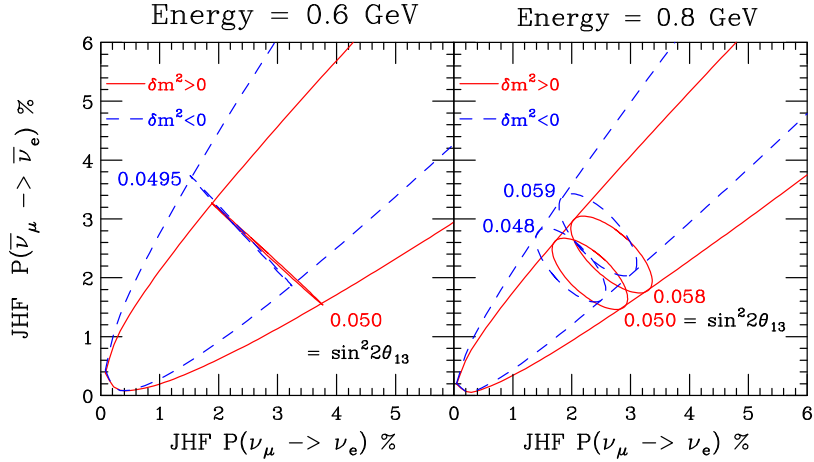}}   
\vglue -0.1cm
\caption{The allowed region in the bi-probability plot at
oscillation maximum (left panel) and at 30\% above 
oscillation maximum (right panel) for $\nu_\mu \rightarrow \nu_e$
verses $\bar{\nu}_\mu \rightarrow \bar{\nu}_e$ for JHF at 295 km.
The ellipses are for fixed $\sin^2 2\theta_{13}$ but allowing
$\delta_{CP}$ to vary from 0 to 2$\pi$.
Except where noted the other mixing parameters are fixed to be 
$|\Delta m^2_{13}| = 2.5 \times 10^{-3}$ eV$^2$,
$\sin^2 2\theta_{23}=1.0$,
$\Delta m^2_{12} = +7 \times 10^{-5}$ eV$^2$
and $\sin^2 2\theta_{12}=0.85$.
The electron density for JHF is fixed to be 
$ Y_e \rho  = 1.15$ g cm$^{-3}$ 
(for NuMI we will use 1.4 g cm$^{-3}$).
\label{13delta}}
\end{figure}

If the probabilities $P(\nu_\mu \rightarrow \nu_e)$ and
$P(\bar{\nu}_\mu \rightarrow \bar{\nu}_e)$ are precisely determined by
long baseline experiments then in general there are four solutions
of parameters $(\theta_{13}, \delta_{CP})$ for a fixed 
value\footnote{\uppercase{T}he ambiguity in $\theta_{23}$ 
will be discussed in the next section.}
of $\theta_{23}$.
This is shown on the right bi-probability diagram (see Ref.
[\refcite{Minakata:2001qm}])
of Fig.~\ref{13delta}
where the four ellipses intersect at a single point.
Two of these ellipse are assuming normal hierarchy and the other two
are the inverted hierarchy.
Note, that the values of $\sin^2 2\theta_{13}$ varies significantly
between the ellipses of the same hierarchy.

The left hand panel of Fig. \ref{13delta} shows that these four
ellipses collapse to a line at the energy that corresponds
to oscillation maximum and that the value of $\sin^2 2\theta_{13}$
can be determined precisely at this special energy, see 
Ref. [\refcite{Kajita:2001sb}].
For a given hierarchy, the complicated  $(\theta_{13}, \delta_{CP})$
degeneracies factorizes into a fixed value for $\theta_{13}$
and a $(\delta_{CP}$, $\pi-\delta_{CP})$ degeneracy.
In general, the hierarchy degeneracy still exists unless nature has 
chosen one of the edges of the allowed region in bi-probability space
represented in Fig.~\ref{13delta}.

\begin{figure}[hbt]
\vglue -0.2cm
\centerline{\epsfxsize=4.3in\epsfbox{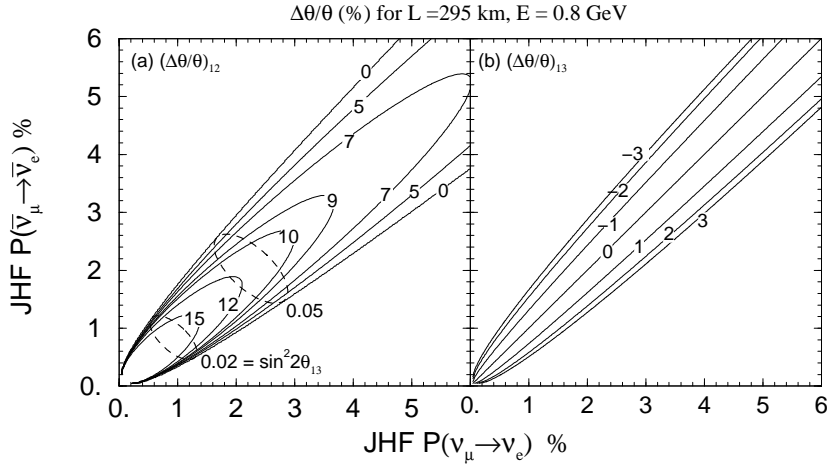}}   
\vglue -0.2cm
\caption{The fractional difference in the values of $\theta_{13}$
for the same hierarchy (left panel) and different hierarchy
(right panel).
\label{dlogtheta}}
\end{figure}

In Fig. \ref{dlogtheta} we have plotted the fractional difference
in the allowed values of $\theta_{13}$ for the same
hierarchy in the left panel and different hierarchies in the right panel.
This fractional uncertainty in the allowed value grows as $\theta_{13}$
gets smaller but for values of  $\theta_{13}$ not too far below
the current Chooz bound of $\sin^2 2\theta_{13} < 0.1$ the fractional
uncertainty is less than say 20\%.\\

\subsection{$\theta_{23}$ Degeneracy}\label{subsec:23delta}
\begin{figure}[bt]
\centerline{\epsfxsize=4.4in\epsfbox{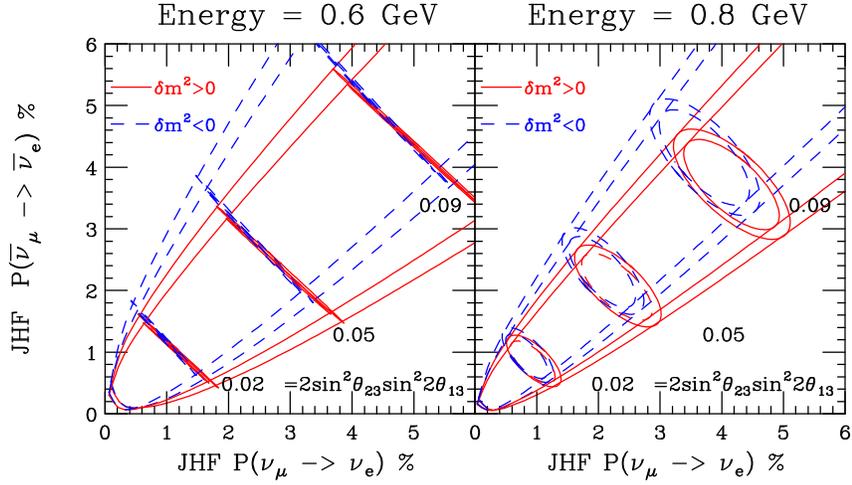}}   
\caption{The bi-probability plots at oscillation maximum (left)
and at an energy 30\% above oscillation maximum (right) assuming that
$\theta_{23}$ differs from $\pi/4$ 
for constant values of 
$(\sin \theta_{23} \sin \theta_{13})$.
Here, $\sin^2 2 \theta_{23} =0.96$ giving 
two solutions for $\sin^2 \theta_{23} = 0.4 ~{\rm and} ~0.6$.
\label{theta23}}
\end{figure}

In general, $\theta_{23}$ is determined from $\nu_\mu \rightarrow \nu_\mu$
disappearance experiments.
Unfortunately in the disappearance probability $\theta_{23}$ appears
as $\sin^2 2\theta_{23}$. If $\sin^2 2\theta_{23}$ differs from
1 by $\epsilon^2$ then the two solutions for
$\sin^2 \theta_{23}$ are $(1\mp\epsilon)/2$. 
Since the appearance probability for $\nu_\mu \rightarrow \nu_e$
depends on $\sin^2 \theta_{23}$ this ambiguity leads an ambiguity
in the determination of $\theta_{13}$. However, the quantity
$(\sin \theta_{23} \sin \theta_{13})$ can be determined accurately
from the appearance experiments. See Fig. \ref{theta23}.

In Fig. \ref{theta23} we have assumed that $\sin^2 2\theta_{23}=0.96=1-(0.2)^2$
and drawn the bi-probability ellipses for 
$(2\sin^2 \theta_{23} \sin^2 2\theta_{13}) 
=0.02, \, 0.05 \, {\rm and} \, 0.09$.
In the right panel the energy is chosen at 30\% above oscillation maximum 
and the four ellipses (two different $\theta_{23}$ times the two hierarchies)
are approximately, but not exactly, degenerate.
The left panel is at oscillation maximum and the degeneracy of the four ellipse
is nearly exact.

\begin{figure}[bt]
\centerline{\epsfxsize=4.4in\epsfbox{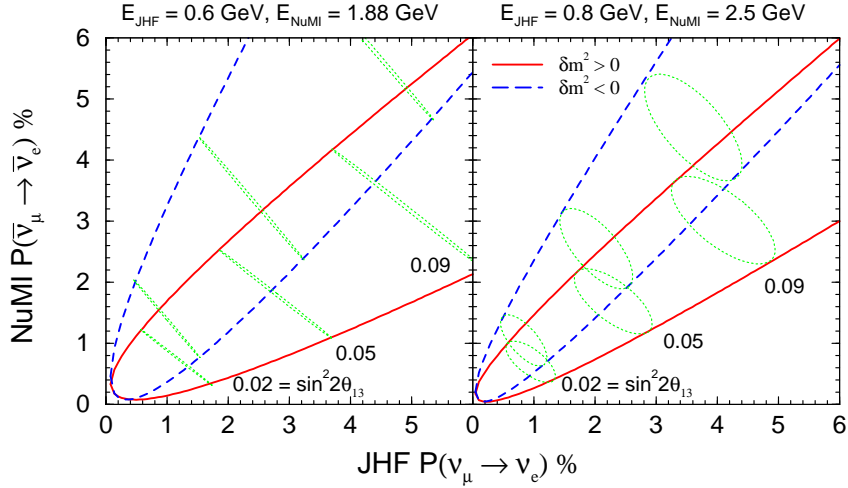}}   
\caption{Allowed region in the bi-probability plane for 
JHF $\nu_\mu \rightarrow \nu_e$
verses NuMI $\bar{\nu}_\mu \rightarrow \bar{\nu}_e$ at 915 km
with represented ellipses for fixed values of  $\sin^2 2 \theta_{13}$.
The left panel is both experiments at oscillation maximum
while the right panel is both experiments 30\% above oscillation
maximum. 
\label{nu-antinu}}
\end{figure}

Thus the ambiguity in the determination of $\theta_{23}$ leads to an ambiguity
in the determination of $\theta_{13}$ in long baseline oscillation
experiments.
However,
the quantity $(\sin \theta_{23} \sin \theta_{13})$ can be precisely
determined especially at oscillation maximum. 
To break the degeneracy in $\theta_{23}$; 
$(\sin \theta_{23} \sin \theta_{13})$ and $(\cos \theta_{23} \sin \delta_{CP})$
can be measured at oscillation maximum then 
$(\cos \theta_{23} \cos \delta_{CP})$ can be determined above oscillation
maximum. The combination of these measurements leads to a determination of
$\cos \theta_{23}$, breaking this $\theta_{23}$ degeneracy.

\begin{figure}[hb]
\vglue 0.11cm
\centerline{\epsfxsize=4.3in\epsfbox{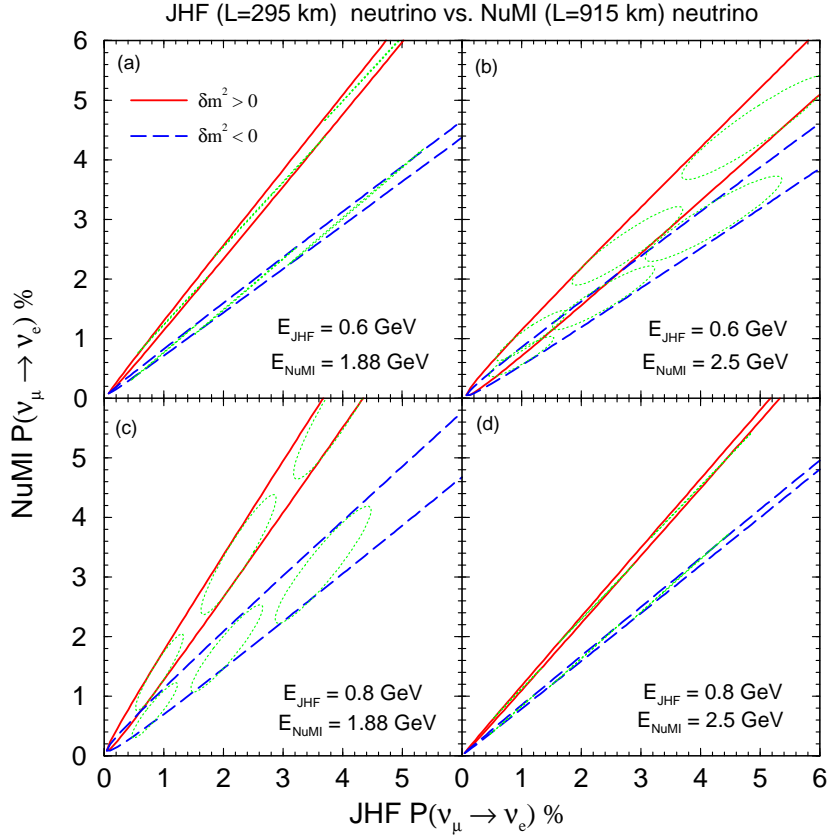}}   
\vglue 0.25cm
\caption{Allowed region in the bi-probability plane for 
JHF $\nu_\mu \rightarrow \nu_e$
verses NuMI $\nu_\mu \rightarrow \nu_e$
with represented ellipses for fixed values of  $\sin^2 2 \theta_{13}$.
The top (left) two panel are JHF (NuMI) at oscillation maximum
while the bottom (right) panel are JHF (NuMI) at 30\% above oscillation
maximum.  The ellipse are for $\sin^2 2\theta_{13}=0.02, \, 0.05 \, {\rm and}
\, 0.09$. 
\label{nu-nu}}
\end{figure}

\vspace{-0.5cm}
\section{JHF/NuMI Complementarity}
Detailed discussions can be found in references
[\refcite{Barger:2002xk}], [\refcite{Huber:2002rs}] and [\refcite{Minakata:2003ca}].
 
\vspace{-0.35cm}
\subsection{Neutrino-Antineutrino}

Fig. \ref{nu-antinu} represents the allowed region in bi-probability
space for one experiment neutrinos and the other anti-neutrinos.
Note the similarity between this plot and the plots where 
both the neutrino and anti-neutrino probability are from the same experiment
(Fig. \ref{13delta} for example).

\subsection{Neutrino-Neutrino}
Fig. \ref{nu-nu} represents the allowed region in bi-probability
space for both experiments neutrinos.  
The allowed regions are narrow ``pencils'' which grow in width 
as the energy of either or both experiments differ from oscillation maximum.
The ratio of slopes of these pencils increases (decreases) as the energy of 
the experiment with smaller (larger) matter effect increases.
For JHF/NuMI this means that the best separation occurs when
\begin{equation}
\frac{E}{L}|_{NuMI}  \le \frac{E}{L}|_{JHF}. 
\end{equation}
The top right panel of Fig. \ref{nu-nu} violates this condition
and there is substantial overlap between the two allowed 
regions, see Ref. [\refcite{Minakata:2003ca}].

\section{Conclusions}
The eight fold parameter degeneracy in $\nu_\mu \rightarrow \nu_e$
can be resolved with multiple measurements in the neutrino and anti-neutrino
channel. A neutrino as well as an anti-neutrino measurement at oscillation
maximum plus a neutrino measurement above oscillation maximum is sufficient
if chosen carefully.  Exploitation of the difference in the matter effect
between JHF and NuMI can be used to determine the mass hierarchy
provided that the E/L of NuMI is smaller than or equal to the E/L of JHF.

\end{document}